\documentclass[12pt]{article}
\usepackage[top=2cm,bottom=3cm,left=2cm,right=2cm]{geometry}
\usepackage{amsmath}
\usepackage{amssymb} 
\usepackage{amsthm}
\usepackage{hyperref}
\usepackage{bm} 

\def\eq#1{(\ref{eq:#1})}
\def\lineup{\!\!\!\!\!\!\!\! &&}

\def\d{\partial}
\def\eps{\epsilon}

\def\LL{{\bf L}}
\def\UU{{\bf U}}
\def\uu{{\bf u}}
\def\QQ{{\bf Q}}
\def\DD{{\bf D}}

\def\SSFT{S_\mathrm{SFT}}

\begin{document}
\begin{titlepage}
\rightline\today

\begin{center}

\vskip 3.5cm

{\large \bf{The closed string field theory action vanishes}}

\vskip 1.0cm

{\large Theodore Erler\footnote{tchovi@gmail.com}}

\vskip 1.0cm

{\it CEICO, FZU - Institute of Physics of the Czech Academy of
Sciences}\\
{\it Na Slovance 2, 182 21 Prague 8, Czech Republic}\\

\vskip 2.0cm

{\bf Abstract}

\end{center}

Using the dilaton theorem, we show that the classical action of closed string field theory vanishes on-shell.

\end{titlepage}

\noindent Consider a field theory with a transformation which changes the normalization of the action. The infinitesimal transformation   satisfies 
\begin{equation}S\Big[\phi^i+\eps f^i[\phi^i]\Big] = (1+\eps)S[\phi^i].\end{equation}
If $\phi^i$ is a solution to the equations of motion, the action is stationary and the left hand side is unchanged. From this it follows that the action vanishes on-shell
\begin{equation}S[\phi^i] = 0,\  \  \text{if } \phi^i  \text{ satisfies equations of motion}.\end{equation}
In this article, we note that classical closed string field theory has a transformation of this form. This follows from the dilaton theorem,  formulated by Bergman and Zwiebach~\cite{BZ}. Therefore the action of closed string field theory vanishes on-shell. Though this observation is rather elementary, it has some importance for understanding closed string tachyon condensation. The value of the on-shell action is also of interest in connection to black hole entropy \cite{Uglum}. Low energy effective actions in string theory also vanish on-shell (at any order in $\alpha'$) \cite{Callan,Tseytlin,Witten}, and the reason is the same.  The action depends on the dilaton field $\phi(x)$ as
\begin{equation}S_\mathrm{eff} = \frac{1}{\kappa^2}\int_M d^D x\, e^{-2\phi(x)} \mathcal{L}\Big[\d\phi(x)\Big],\label{eq:effS}\end{equation}
where $\kappa$ is the coupling constant and $\mathcal{L}$ depends on derivatives of $\phi(x)$ and additionally the metric and other massless fields. Here a constant shift of the dilaton changes the normalization of the action, and therefore the action vanishes on-shell. It is important to qualify that this assumes that we are working on a compact target space. For non-compact target spaces, the action only needs to be stationary with respect to variations which vanish at infinity. A constant shift of the dilaton is not a variation of this form, and for this reason the on-shell action on non-compact target spaces can receive nonzero boundary contributions. We assume target space is compact, and comment on the noncompact case in the end.

The goal in what follows is to reexpress the closed string field theory action in a form which is proportional to the equations of motion. Then it is manifest that the action vanishes on-shell. We will employ the coalgebra description of $L_\infty$ algebras, reviewed for example in \cite{ClosedSFT_Erler}. The classical action of closed string field theory takes the form
\begin{equation}
\SSFT(\kappa)  = \frac{1}{2!}\omega(\Phi,Q\Phi)  +  \frac{\kappa}{3!}\omega(\Phi,L_2(\Phi,\Phi))+\frac{\kappa^2}{4!}\omega(\Phi,L_3(\Phi,\Phi,\Phi))+...\ ,
\end{equation}
where $\Phi$ is the dynamical closed string field. $\Phi$ carries ghost number 2, is Grassmann even, and is subject to $b_0^-$ and level matching constraints. The object $\omega$ is the symplectic form, and $L_n$ are a sequence of Grassmann odd string products. The products are {\it cyclic} in that they satisfy
\begin{equation}
\omega(A_1,L_n(A_2,...,A_{n+1}))  = -(-1)^{|A_1|}\omega(L_n(A_1,...,A_n),A_{n+1}).\\
\end{equation}
In addition the string products are graded symmetric and satisfy $L_\infty$ relations. This may be expressed compactly in the coalgebra formalism by introducing a coderivation on the symmetrized tensor algebra
\begin{equation}
\LL(\kappa) = \QQ+\kappa \LL_2+\kappa^2\LL_3+...\ .
\end{equation}
The $L_\infty$ relations are equivalent to the statement that this is nilpotent: $\LL(\kappa)^2=0$.

\vspace{.25cm}

The classical part of Bergman and Zwiebach's result \cite{BZ} is the existence of a field variation $\delta_U\Phi$ satisfying
\begin{equation}\delta_U \SSFT(\kappa) = \kappa\frac{d}{d\kappa}\SSFT(\kappa).\end{equation}
The infinitesimal field redefinition $\Phi\to\Phi+\eps \delta_U\Phi$ adjusts the coupling constant from $\kappa$ to $\kappa(1+\eps)$. The variation can be expressed as an expansion in powers of the string field 
\begin{equation}\delta_U\Phi =\frac{1}{\kappa}D +U_1(\Phi) +\frac{\kappa}{2!}U_2(\Phi,\Phi)+\frac{\kappa^2}{3!}U_3(\Phi,\Phi,\Phi)+...\ .\label{eq:fieldred}\end{equation}
The  leading term is a shift of the vacuum given by the ghost-dilaton state
\begin{equation} D = (c_1c_{-1}-\overline{c}_1\overline{c}_{-1})|0\rangle.\end{equation}
The next term takes the form
\begin{equation}U_1(\Phi) = -L_2(\chi,\Phi),\ \ \ \chi = -\frac{1}{2}(c_0-\overline{c}_0)|0\rangle.\end{equation}
The remaining terms implement a deformation of the vertices expanded around the ghost dilaton state into vertices with insertions of $D$ and $\chi$ in a form where the correlation functions can be evaluated to produce the Euler number of the surface represented by the vertex, as originally envisaged by Distler and Nelson \cite{DN1,DN2}. The variation is defined by a sequence $D,U_1,U_2,U_3...$ of Grassmann even, graded symmetric and cyclic string products which can be  packaged into a coderivation
\begin{equation}\UU(\kappa) =  \frac{1}{\kappa}\DD+\UU_1+\kappa\UU_2+\kappa^2\UU_3+...\ .\end{equation}
In coalgebra language, the variation may be expressed as  
\begin{equation}\delta_U\Phi = \pi_1\UU(\kappa)e^{\Phi},\end{equation}
where $\pi_1$ is the projection onto the 1-string component of the symmetrized tensor algebra and $e^\Phi$ is the group-like element of the symmetrized tensor algebra generated by $\Phi$. The fact that $\delta_U$ adjusts the coupling constant is guaranteed by certain algebraic relations satisfied by the string products, which are encoded in the differential equation 
\begin{equation}
\kappa\frac{d}{d\kappa}\LL(\kappa)  = [\LL(\kappa),\UU(\kappa)].\label{eq:coalgeq}
\end{equation}
An analogous field variation should exist in closed superstring field theories, though there does not appear to be an explicit discussion of it. This  question is closely related to the proof of background independence under marginal deformation of the worldsheet CFT \cite{S,SZ1,SZ2}, which was discussed for the superstring in \cite{Ssuper}.

\vspace{.25cm}

The field variation \eq{fieldred} is not exactly what we need since it does not change the normalization of the action. Rather, it adjusts the coupling constant as it appears in powers multiplying the string vertices. This however can be fixed by redefining the string field
\begin{equation}\Phi\to \frac{1}{\kappa}\Phi\end{equation}
so that the coupling constant appears only as an overall $1/\kappa^2$ multiplying the action. However, it will be necessary to redefine the string field in a manner which preserves the symplectic form. Simply scaling the closed string vector space by $1/\kappa$ does not preserve the symplectic form. For this reason we introduce a shifted ghost number operator
\begin{equation}\Gamma \equiv \text{gh}-\frac{5}{2}.\end{equation}
The closed string field at ghost number $2$ has shifted ghost number $-1/2$. This operator preserves the closed string symplectic structure, i.e. it is cyclic
\begin{equation}\omega(\Gamma A,B) + \omega(A,\Gamma B) = 0.\end{equation}
If $\bm{\Gamma}$ is the coderivation corresponding to $\Gamma$, we have the properties
\begin{eqnarray}
[\bm{\Gamma},\LL_n] \lineup = \frac{n+1}{2}\LL_n,\\
\ [\bm{\Gamma},\UU_n]\lineup = \frac{n-1}{2}\UU_n.
\end{eqnarray}
From this we learn that
\begin{eqnarray}
\frac{1}{\kappa^2}\LL \lineup = \kappa^{-2\bm{\Gamma}}\LL(\kappa)\kappa^{2\bm{\Gamma}},\label{eq:LLk}\\
\UU \lineup = \kappa^{-2\bm{\Gamma}}\UU(\kappa)\kappa^{2\bm{\Gamma}}.\label{eq:UUk}
\end{eqnarray}
where $\LL$ and $\UU$ are $\LL(\kappa)$ and $\UU(\kappa)$ evaluated at $\kappa=1$. Therefore we can accomplish the desired result by transforming the string field as $\Phi \to \kappa^{2\Gamma}\Phi$, and the symplectic structure is preserved. Since the closed string field has ghost  number $2$, this reduces to the transformation $\Phi \to \kappa^{-1}\Phi$ as expected. The differential equation \eq{coalgeq} for $\LL(\kappa)$ implies a useful relation for $\LL$. To find it, we take the derivative of \eq{LLk}. On the left hand side, this gives simply
\begin{equation}
\kappa\frac{d}{d\kappa}\left(\frac{1}{\kappa^2}\LL \right)= -2\frac{1}{\kappa^2}\LL.
\end{equation}
On the right hand side, using \eq{coalgeq} and \eq{UUk} gives
\begin{equation}
\kappa\frac{d}{d\kappa}\Big(\kappa^{-2\bm{\Gamma}}\LL(\kappa)\kappa^{2\bm{\Gamma}}\Big) = \frac{1}{\kappa^2}[\LL,2\bm{\Gamma}+\UU].
\end{equation}
Introducing the coderivation
\begin{equation}\uu = -\bm{\Gamma}-\frac{1}{2}\UU,
\end{equation}
we find
\begin{equation}
\LL = [\LL,\uu].\label{eq:LLu}
\end{equation}
This relation implies that the field variation defined by 
\begin{equation}\delta_u\Phi = \pi_1 \uu e^\Phi\end{equation}
satisfies
\begin{equation}\delta_u \SSFT = \SSFT,\end{equation}
where $\SSFT$ is the action $\SSFT(\kappa)$ evaluated at $\kappa=1$. Therefore the transformation $\delta_u\Phi$ changes the normalization of the action, as desired. 

\vspace{.25cm}

Next we use \eq{LLu} to derive an expression for the action involving the products of both $\LL$ and $\uu$. We normalize $\Phi$ so that the  coupling constant appears as a multiplicative factor of $1/\kappa^2$ in front of the action. As in \cite{ClosedSFT_Erler} we may express the action in the coalgebra formalism  as 
\begin{equation}\frac{1}{\kappa^2}\SSFT = \frac{1}{\kappa^2}\int_0^1dt\,\omega\Big(\dot{\Phi}(t),\pi_1 \LL e^{\Phi(t)}\Big).\label{eq:S1}\end{equation}
where $\Phi(t)$ is a function of the closed string field $\Phi$ and the parameter $t$ subject to the boundary conditions
\begin{equation}\Phi(1)=\Phi,\ \ \ \Phi(0)=0,\end{equation}
and $\dot{\Phi}(t)$ indicates differentiation with respect to $t$. Substituting \eq{LLu} gives 
\begin{eqnarray}
\frac{1}{\kappa^2} \SSFT \lineup = \frac{1}{\kappa^2}\int_0^1dt\,\omega\Big(\dot{\Phi}(t),\pi_1[\LL,\uu]e^{\Phi(t)}\Big)\nonumber\\
\lineup = \frac{1}{\kappa^2}\int_0^1dt\,\omega\bigg(\dot{\Phi}(t),\pi_1\LL\Big(\Big(\pi_1\uu e^{\Phi(t)}\Big)\wedge e^{\Phi(t)}\Big)\bigg)-\frac{1}{\kappa^2}\int_0^1dt\,\omega\bigg(\dot{\Phi}(t),\pi_1\uu\Big(\Big(\pi_1\LL e^{\Phi(t)}\Big)\wedge e^{\Phi(t)}\Big)\bigg).
\nonumber\\
\end{eqnarray}
Next we observe that the products of $\LL$ and $\uu$ are cyclic. This allows us to shift the coderivations from the second entry of the symplectic form to the first
\begin{eqnarray}
\frac{1}{\kappa^2}\SSFT \lineup = -\frac{1}{\kappa^2}\int_0^1dt\,\omega\bigg(\pi_1\LL\Big(\dot{\Phi}(t)\wedge e^{\Phi(t)}\Big),\pi_1\uu e^{\Phi(t)}\bigg)+
\frac{1}{\kappa^2}\int_0^1dt\,\omega\bigg(\pi_1\uu\Big(\dot{\Phi}(t)\wedge e^{\Phi(t)}\Big),\pi_1\LL e^{\Phi(t)}\bigg).
\nonumber\\
\end{eqnarray}
Using antisymmetry of the symplectic form this may be written as an integral of a total derivative with respect to $t$. Therefore the action is expressed
\begin{eqnarray}
\frac{1}{\kappa^2}\SSFT \lineup = \frac{1}{\kappa^2}\omega\Big(\pi_1\uu e^{\Phi},\pi_1\LL e^{\Phi}\Big).\label{eq:result}
\end{eqnarray}
The classical equations of motion of closed string field theory are
\begin{equation}
\pi_1\LL e^{\Phi} = 0,
\end{equation}
from which it follows that the action vanishes on-shell. 

\vspace{.25cm}

One application where closed string field theory is likely more powerful than the $\alpha'$ expansion of effective field theory is in describing the physics of tachyon condensation. In this connection, the vanishing of the action is counterintuitive.  When a closed string tachyon condenses, it is hard to imagine that it is not searching for a minimum of the effective  potential. But apparently there is no vacuum which can bring the potential any lower, at least on compact target spaces. Still, a candidate for  the endpoint of bulk closed string tachyon condensation of the bosonic string was found by Yang and Zwiebach~\cite{YZ} using level truncation and keeping vertices up to quartic order. The solution was argued to have vanishing action based on the characteristic dilaton dependence of the low energy effective field theory, and it would now seem this expectation is on solid footing. Despite some efforts to improve initial results \cite{YM,Moeller}, the existence of the Yang-Zwiebach vacuum has not yet been fully established.

\vspace{.25cm}

A perhaps more reliable class of solutions concern the condensation of twisted tachyons on $\mathbb{C}/\mathbb{Z}_n$ orbifolds \cite{ASP}, expected to achieve a decrease in deficit angle at the conical singularity. This was studied in closed bosonic string field theory by Okawa and Zwiebach~\cite{OZ}. Truncating the action to include twisted tachyons and low momentum components of the bulk tachyon, the action was found to be nonzero and in reasonable agreement with an approximate conjecture due to Dabholkar \cite{D}. However, Dabholkar's conjecture is unlikely to be correct once massless states are accounted for. Presently it is unclear what the exact value of the action should be. Perhaps it will vanish, but there is reason to doubt it. The orbifold is noncompact, and condensation changes the asymptotic structure of spacetime. Presumably if the action is nonzero it is the result of total derivative contributions which were implicitly dropped in the derivation of~\eq{result}.

\vspace{.25cm}

However, even if the depth of the potential for twisted tachyons could be computed to high accuracy, it is questionable whether it would be correct. The reason is that the present formulation of closed string field theory is not complete for backgrounds with nonvanishing action. This can be seen as follows. The action can be decomposed into a field independent constant term $S_0$ and a term $S_\text{hom}$  which is homogeneous in the string field
\begin{equation}\frac{1}{\kappa^2}\SSFT=\frac{1}{\kappa^2}\Big(S_0 + S_\mathrm{hom}\Big).\end{equation}
The constant piece ensures that the action is nonzero when the string field vanishes. Naively, we can take $S_\mathrm{hom}$ to be the usual closed string field theory action \eq{S1} formulated on a conformal background which also happens to have nonvanishing action. The problem with this is that the dilaton shift must adjust the coupling constant in front of the constant term. This implies that the variation $\delta_u S_\mathrm{hom}$ must have an inhomogeneous term
\begin{equation}\delta_u S_\mathrm{hom} = S_0 + S_\mathrm{hom}.\end{equation}
Even accounting for subtleties with integration by parts or possible modifications of $\delta_u$, such a relation is not possible with the ordinary closed string field theory action. At the very least the action must be supplemented by a tadpole vertex. This is surprising since the background is conformal and the ordinary closed string field theory action is naively gauge invariant on its own. To see what is happening it is helpful to look at the low energy effective action. Let $\phi_0(x)$ be the dilaton part of a classical solution of the effective field theory which leads to a nonvanishing action. We can expand the effective action \eq{effS} around this solution by writing $\phi(x)=\phi_0(x)+\varphi(x)$, where $\varphi(x)$ is the dilaton fluctuation field. We do not expect to find a tadpole vertex for the fluctuation field because $\phi_0(x)$ satisfies the equations of motion. But in actuality demonstrating this requires integration by parts, which produces a boundary term
\begin{equation}
\frac{1}{\kappa^2} \int_{\d M} d^{D-1} x\,  e^{-2\phi_0(x)} \frac{\d\mathcal{L}[\d\phi_0(x)]}{\d(\d_\perp\phi_0(x))}\varphi(x)
\end{equation}
The boundary term is a tadpole vertex and is precisely what is needed to account for inhomogeneous contribution to the dilaton shift of the action. Therefore closed string field theory on backgrounds with nonzero action must be refined by some kind of boundary term. Perhaps a complete description would account for the Gibbons-Hawking-York term \cite{GHY} and other boundary terms which must be added to the low energy effective action.  It would be very interesting to understand what boundary terms look like in string field theory.

\vspace{.25cm}

One can also try to understand the value of the action from the point of view of first quantized string theory. The on-shell action is believed to be related to the vacuum sphere amplitude in the background. Computing such amplitudes requires cancellation between the infinite volume of the conformal Killing group and the infinite volume of target space, and typically one does not know how to extract a well-defined result. For vacuum disk amplitudes, the volume of the conformal Killing group can be assigned a finite negative value after subtracting an unphysical divergence \cite{Liu,Eberhart}, and one can consistently associate such amplitudes with D-brane tensions. A similar procedure does not appear to apply to vacuum sphere amplitudes.  Recently, the vacuum sphere amplitude has been evaluated in a semiclassical limit of Liouville theory \cite{Stanford} and was confirmed to be nonzero and in agreement with the matrix integral result. There is also a possibility that vacuum sphere amplitudes could be computed through a novel gauge fixing of the Polyakov path integral, along the lines of the computation of the 2-point amplitude in \cite{Maldacena}. These considerations may inform how we should deal with backgrounds with nonvanishing action in string field theory.

\vspace{.25cm}

Therefore the value of the on-shell action in closed string field theory remains unclear in most interesting cases. But the fact that it vanishes up to boundary terms is an important baseline for consideration. 

\vspace{.5cm}

\noindent{\bf Acknowledgements}

\vspace{.25cm}

\noindent The author thanks A. Sen, B. Zwiebach and H. Erbin for correspondence, and especially R. Mahajan for discussions which clarified the broader context of this problem. The author also thanks the organizers of the SFT journal club, where the author learned about Mahajan's interesting work. This work is supported by European Structural and Investment Fund and the Czech Ministry of Education, Youth and Sports (Project CoGraDS- CZ.02.1.01/0.0/0.0/15\_ 003/0000437).


\begin{thebibliography}{99}


\bibitem{BZ}

O.~Bergman and B.~Zwiebach,
``The Dilaton theorem and closed string backgrounds,''
Nucl. Phys. B \textbf{441}, 76-118 (1995)
[arXiv:hep-th/9411047 [hep-th]].

\bibitem{Uglum}

L.~Susskind and J.~Uglum,
``Black hole entropy in canonical quantum gravity and superstring theory,''
Phys. Rev. D \textbf{50}, 2700-2711 (1994)
[arXiv:hep-th/9401070 [hep-th]].


\bibitem{Callan}

C.~G.~Callan, Jr., I.~R.~Klebanov and M.~J.~Perry,
``String Theory Effective Actions,''
Nucl. Phys. B \textbf{278}, 78-90 (1986)

\bibitem{Tseytlin}

A.~A.~Tseytlin,
``Mobius Infinity Subtraction and Effective Action in $\sigma$ Model Approach to Closed String Theory,''
Phys. Lett. B \textbf{208}, 221-227 (1988)


\bibitem{Witten}

Y.~Chen, J.~Maldacena and E.~Witten,
``On the black hole/string transition,''
[arXiv:2109.08563 [hep-th]].

\bibitem{ClosedSFT_Erler}

T.~Erler,
``Four Lectures on Closed String Field Theory,''
Phys. Rept. \textbf{851}, 1-36 (2020)
[arXiv:1905.06785 [hep-th]].

\bibitem{DN1}

J.~Distler and P.~C.~Nelson,
``Topological couplings and contact terms in 2-d field theory,''
Commun. Math. Phys. \textbf{138}, 273-290 (1991)

\bibitem{DN2}

J.~Distler and P.~C.~Nelson,
``The Dilaton equation in semirigid string theory,''
Nucl. Phys. B \textbf{366}, 255-272 (1991)

\bibitem{S}

A.~Sen,
``On the Background Independence of String Field Theory,''
Nucl. Phys. B \textbf{345}, 551-583 (1990)

\bibitem{SZ1}

A.~Sen and B.~Zwiebach,
``A Proof of local background independence of classical closed string field theory,''
Nucl. Phys. B \textbf{414}, 649-714 (1994)
[arXiv:hep-th/9307088 [hep-th]].


\bibitem{SZ2}

A.~Sen and B.~Zwiebach,
``Quantum background independence of closed string field theory,''
Nucl. Phys. B \textbf{423}, 580-630 (1994)
[arXiv:hep-th/9311009 [hep-th]].


\bibitem{Ssuper}

A.~Sen,
``Background Independence of Closed Superstring Field Theory,''
JHEP \textbf{02}, 155 (2018)
[arXiv:1711.08468 [hep-th]].

\bibitem{YZ}

H.~Yang and B.~Zwiebach,
``A Closed string tachyon vacuum?,''
JHEP \textbf{09}, 054 (2005)
[arXiv:hep-th/0506077 [hep-th]].

\bibitem{YM}

N.~Moeller and H.~Yang,
``The Nonperturbative closed string tachyon vacuum to high level,''
JHEP \textbf{04}, 009 (2007)
[arXiv:hep-th/0609208 [hep-th]].

\bibitem{Moeller}

N.~Moeller,
``Closed Bosonic String Field Theory at Quintic Order. II. Marginal Deformations and Effective Potential,''
JHEP \textbf{09}, 118 (2007)
[arXiv:0705.2102 [hep-th]].



\bibitem{ASP}

A.~Adams, J.~Polchinski and E.~Silverstein,
``Don't panic! Closed string tachyons in ALE space-times,''
JHEP \textbf{10}, 029 (2001)
[arXiv:hep-th/0108075 [hep-th]].

\bibitem{OZ}

Y.~Okawa and B.~Zwiebach,
``Twisted tachyon condensation in closed string field theory,''
JHEP \textbf{03}, 056 (2004)
[arXiv:hep-th/0403051 [hep-th]].

\bibitem{D}

A.~Dabholkar,
``Tachyon condensation and black hole entropy,''
Phys. Rev. Lett. \textbf{88}, 091301 (2002)
[arXiv:hep-th/0111004 [hep-th]].


\bibitem{GHY}

G.~W.~Gibbons and S.~W.~Hawking,
``Action Integrals and Partition Functions in Quantum Gravity,''
Phys. Rev. D \textbf{15}, 2752-2756 (1977)



\bibitem{Liu} 

J.~Liu and J.~Polchinski,
``Renormalization of the Mobius Volume,''
Phys. Lett. B \textbf{203}, 39-43 (1988)

\bibitem{Eberhart}

L.~Eberhardt and S.~Pal,
``The disk partition function in string theory,''
JHEP \textbf{08}, 026 (2021)
[arXiv:2105.08726 [hep-th]].

\bibitem{Stanford}

R.~Mahajan, D.~Stanford and C.~Yan,
``Sphere and disk partition functions in Liouville and in matrix integrals,''
[arXiv:2107.01172 [hep-th]].


\bibitem{Maldacena}

H.~Erbin, J.~Maldacena and D.~Skliros,
``Two-Point String Amplitudes,''
JHEP \textbf{07}, 139 (2019)
[arXiv:1906.06051 [hep-th]].


\end{thebibliography}
\end{document}